**Physics-data-driven intelligent optimization for large-scale meta-devices**

Yingli Ha[1,2,†], Yu Luo[1,2,†], Mingbo Pu[1,2,3,*], Fei Zhang[1,2], Qiong He[1], Jinjin Jin[1], Mingfeng Xu[1,2,3], Yinghui Guo[1,2,3], Xiaogang Li[4], Xiong Li[1,3], Xiaoliang Ma[1,3], and Xiangang Luo[1,3,*]

[1] State Key Laboratory of Optical Technologies on Nano-Fabrication and Micro-Engineering, Institute of Optics and Electronics, Chinese Academy of Sciences, Chengdu 610209, China.
[2] Research Center on Vector Optical Fields, Institute of Optics and Electronics, Chinese Academy of Sciences, Chengdu 610209, China.
[3] School of Optoelectronics, University of Chinese Academy of Sciences, Beijing 100049, China.
[4] Tianfu Xinglong Lake Laboratory, Chengdu 610299, China.

[†] These authors contributed equally: Yingli Ha, Yu Luo.
* Correspondence: pmb@ioe.ac.cn ; lxg@ioe.ac.cn

**Abstract**

Meta-devices have gained significant attention and have been widely utilized in optical systems for focusing and imaging, owing to their lightweight, high-integration, and exceptional-flexibility capabilities. However, based on the assumption of local phase approximation, traditional design method neglect the local lattice coupling effect between adjacent meta-atoms, thus harming the practical performance of meta-devices. Using physics-driven or data-driven optimization algorithms can effectively solve the aforementioned problems. Nevertheless, both of the methods either involve considerable time costs or require a substantial amount of data sets. Here, we propose a physics-data-driven approach based "intelligent optimizer" that enables us to adaptively modify the sizes of the studied meta-atom according to the sizes of its surrounding ones. Such a scheme allows to mitigate the undesired local lattice coupling effect, and the proposed network model works well on thousands of datasets with a validation loss of $3\times10^{-3}$. Experimental results show that the 1-mm-diameter metalens designed with the "intelligent optimizer" possesses a relative focusing efficiency of 93.4% (as compared to ideal focusing) and a Strehl ratio of 0.94. In contrast to the previous inverse design method, our method significantly boosts designing efficiency with five orders of magnitude reduction in time. Our design approach may sets a new paradigm for devising large-scale meta-devices.

**1 Introduction**

In the past decade, meta-devices have shown remarkable advantages of lightweight, high integration, and high flexibility capabilities [1-6], which have led to the emergence of many new type of flat optical components, including lenses [7, 8], holograms [9, 10], vector beam generator [2, 11], and optical encryption [12, 13]. Traditional design method of meta-devices is based on phase matching, which assigns a pre-defined phase profile to each meta-atom [14]. The method is based on the assumption of local phase approximation and each phase of the meta-atom is obtained using periodic boundary conditions. Therefore, for a meta-devices with aperiodic layouts, traditional design method neglects the local lattice coupling effect between adjacent meta-atoms, thus leading to errors between the real electric field distribution and the ideal one [15]. Optimization algorithms can be used to improve the efficiency of the meta-device, but it is still not possible to completely eliminate the error introduced by coupling effects through independent optimization of each pixel [16]. Therefore, it is necessary to consider the interaction of the entire device and optimize it as a whole to obtain high-performance meta-devices.

Inverse design methods could iteratively update the real vectorial electric field for each pixel, gradually approaching the ideal electric field distribution, and can effectively solve the problems of the traditional design method [17-19]. Based on physics-driven or data-driven optimization algorithms, abundant methods of inverse design techniques have been studied [20-24]. Among them, adjoint-based optimization has demonstrated success in realizing device functions, such as metalens [25-27], disordered metasurface [19], and deflector [28, 29]. The global optimal solution could be achieved by searching a large number of initial structures [18, 30]. However, the above physics-driven methods are usually used to optimize small-scale meta-devices, typically, <100$\lambda$ [25]. Most recently, more researchers have focused on inverse design methods based on the deep learning (DL) network [31-35], and considerable results have been achieved [36-40], such as deflector [41, 42], 3D vectorial holography [43], multi-function meta-device [44], and multi-band absorber [37]. In addition, researchers try to alleviate the burden of numerical calculations, accelerate optimization, by using semi-supervised learning strategies [45], self-supervised learning strategies [46], data-driven evolutionary algorithms [47], and physics-driven DL methods [41, 42, 48]. However, further optimization of the inverse design is highly challenging for large-scale meta-devices with aperiodic meta-atoms. One current feasible design method for large-scale aperiodic meta-devices is based on adjoint simulation and Chebyshev interpolation. Although this method can improve the efficiency of the meta-device, it still takes several hours to optimize a device with the diameter of $1000\lambda$. In addition, most effective data-driven methods (DL methods) rely heavily on vast amounts of data sets (tens of thousands of data) [49], complex network models (dozens of layers) [50], and poor generalization capabilities [41, 42, 48].

In this work, we proposed a physics-data-driven design method which employs a strategy of collaboration between multi-objective optimization algorithm and DL method. We have developed an "intelligent optimizer" that utilizes an end-to-end design framework to achieve optimized design of large-scale meta-devices. Compared with physics-driven or data-driven optimization methods, our method has the advantages of simple network models (16 layers of neurons), low data set requirements (~5000 data sets), and short optimization time (11min20s@1mm$^2$). The simulation results provide evidence that the proposed method is capable of effectively optimizing the polarization-multiplexed metalens. In principle, the size of the designed metalens can be as large as hundreds of millimeters. It is experimentally demonstrated that a 1-mm-diameter metalens with an F-number of 1 (NA=0.44) could achieve a relative focusing efficiency of 93.4%. Specifically, the same network is also applicable to optimize metalenses with higher F-numbers. To illustrate the advantages of the proposed method, **Table 1** is listed which sums up the representative parameters of various literature. In Table 1, RGB means the working wavelengths ($\lambda$) are red (470 nm), green (532 nm), and blue (633 nm).

**Table 1.** Examples to show the representative parameters and performance of various methods.

| Ref. | $\lambda$ (nm) | Material | NA | D (μm) | Dim-ension | Efficiency | Method | Time |
|---|---|---|---|---|---|---|---|---|
| Cai et al.[21] | 532 | TiO$_2$ | 0.51 | 24 | 1D | 60% | genetic algorithm | 1000s |
| Liang et al.[16] | 532 | TiO$_2$ | 0.98 | - | 2D | 67% | hybrid optimization algorithm | - |
| Phan et al.[51] | 640 | SOI | 0.5 | 200 | 1D | 89% | topology optimization | 100 hours |
| Mansouree et al.[25] | 850 | a-Si | 0.75 | 52 | 2D | 65% | adjoint optimization | 97 min |
| Pestourie et al.[52] | RGB | TiO$_2$ | 0.3 | 235 | 1D | - | "locally periodic" approximation | 250 s |
| Li et al.[27] | RGB | TiO$_2$ | 0.7 | 1000 $\lambda$ | 2D | 15% | Conservative convex separable approximation | few hours |
| An et al.[32] | 1550 | p-Si | 0.72 | 32 | 1D | 77.62% | Deep learning | 200s |
| Arbabi et al.[53] | 1550 | a-Si | 0.37 | 50 | 2D | 82% | high-contrast gratings | - |
| **This work** | **1550** | **SOS** | **0.44** | **50** | **2D** | **95.7%** | **physics-data-driven method** | **15s** |
| **This work** | **1550** | **SOS** | **0.44** | **1000** | **2D** | **93.4%** | **physics-data-driven method** | **11 min20s** |

**2 Design Principle**

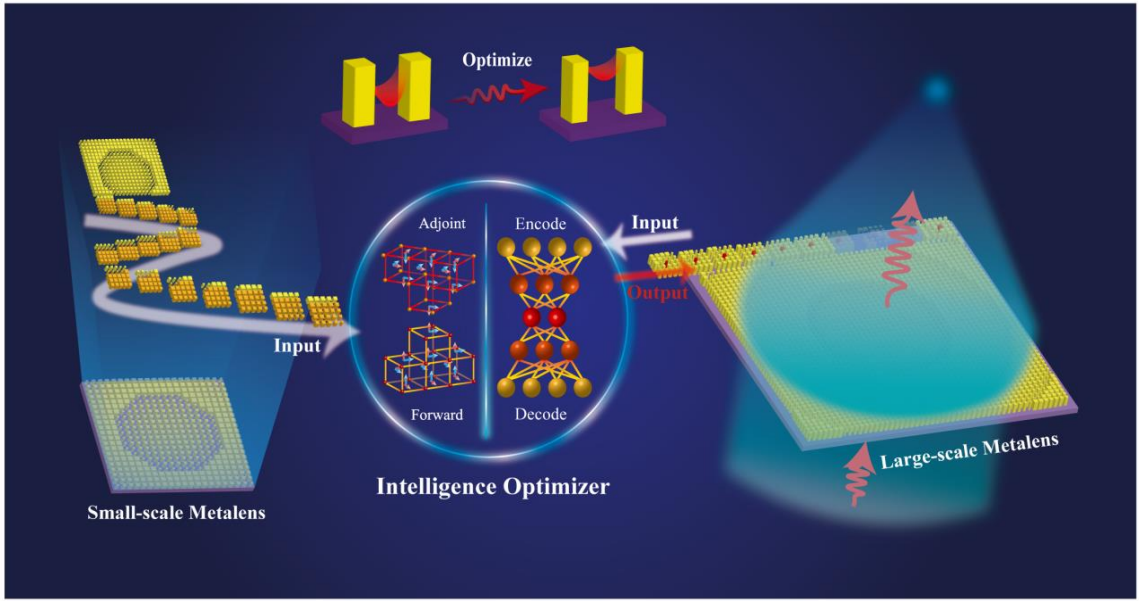

**Figure 1** Working principle of the "intelligent optimizer". The "intelligent optimizer" has been designed based on Adj and DL methods. Data sets of the DL network are obtained from the small-scale metalens optimized by the Adj method. Super meta-atoms of large-scale metalens are fed into the network one by one. Output meta-atoms are spliced together to create a new metalens with a higher focusing efficiency.

Compared with previous DL networks that learn the performance of the full-device, the proposed network is based on an end-to-end design. As shown in **Figure 1**, the "intelligent optimizer" consists of two optimization methods: the physics-driven method (adjoint-based shape optimization, Adj method) and the data-driven method (DL method). The adjoint-based shape optimization is utilized to produce high-quality data sets, while the DL method is employed to enhance the focusing efficiency of the large-scale metalens. To describe the target of the optimization algorithm more intuitively, the catenary-shaped electric field lines between two adjacent meta-atoms are used to indicate the coupling efficiency between the two adjacent meta-atoms [54, 55]. The inputs for the network are geometric parameters of a super meta-atom (composed of adjacent $N \times N$ meta-atoms) which are sliced from the initial metalens in a specific order. The outputs are geometric parameters of one meta-atom which is sliced from the optimized metalens. The new metalens can be obtained by splicing the optimized meta-atoms in a consistent sequence. As a result, the new metalens has a significantly higher focusing efficiency than the initial one.

High-quality data sets can provide sufficient and accurate information to the network, allowing it to capture features better, improve its generalization ability, and increase validation accuracy. Therefore, choosing appropriate and high-quality data sets is an indispensable step for training DL networks. Here, metalenses are artificially used as the basis of data sets, which will be helpful in constructing high-quality data sets. Essentially, the optimization of the metalens is to make the real phase distribution closer to the ideal one which can be obtained by Eq.1. Therefore, coupling effects among random phase arrangement structures are neglected in this study. This strategy has contributed to reducing the amount of data set requirements and the complexity of the model and allows the network to converge faster. Furthermore, to endow our network with optimization capabilities, we will use initial and optimized metalenses as inputs and outputs. The optimization algorithm is the Adj method and more details pertaining to the Adj method are provided in Supporting Information, Sections 1.

Due to the long calculation time and the large memory capacity resulting from many iterations of simulations, the diameters of metalenses are set as 30.5 μm and 40.5 μm, and the mesh size of the electrical field simulation is set as 50 nm. The optimization of the small-scale metalens can be separated into 3 steps. First, the initial meta-atoms for the initial metalens are selected. Eight kinds of meta-atoms cover the phase range 0-to-2π by replacing the ideal phase distribution with an approximated step-function. To maintain high transmission efficiency, the period and height of meta-atoms are set as 500 nm and 1000 nm, respectively. It is important to note that the period and height of the metalens optimized by the DL method should be consistent with the small-scale metalens. Second, initial metalenses are designed. We designed polarization-multiplexed metalenses that can independently control *x*- and *y*- polarized light. The *F* numbers of the above metalenses are both 1, and the focus spots for *x*-polarized incident light and *y*-polarized incident light are at the center of the focal plane. Therefore, the modulation of *x*- and *y*-polarization is identical for each initial metalens, leading to the same width and length for each meta-atom. The width/length of each meta-atom for the initial metalens is determined according to the phase profile given by Eq. 1.

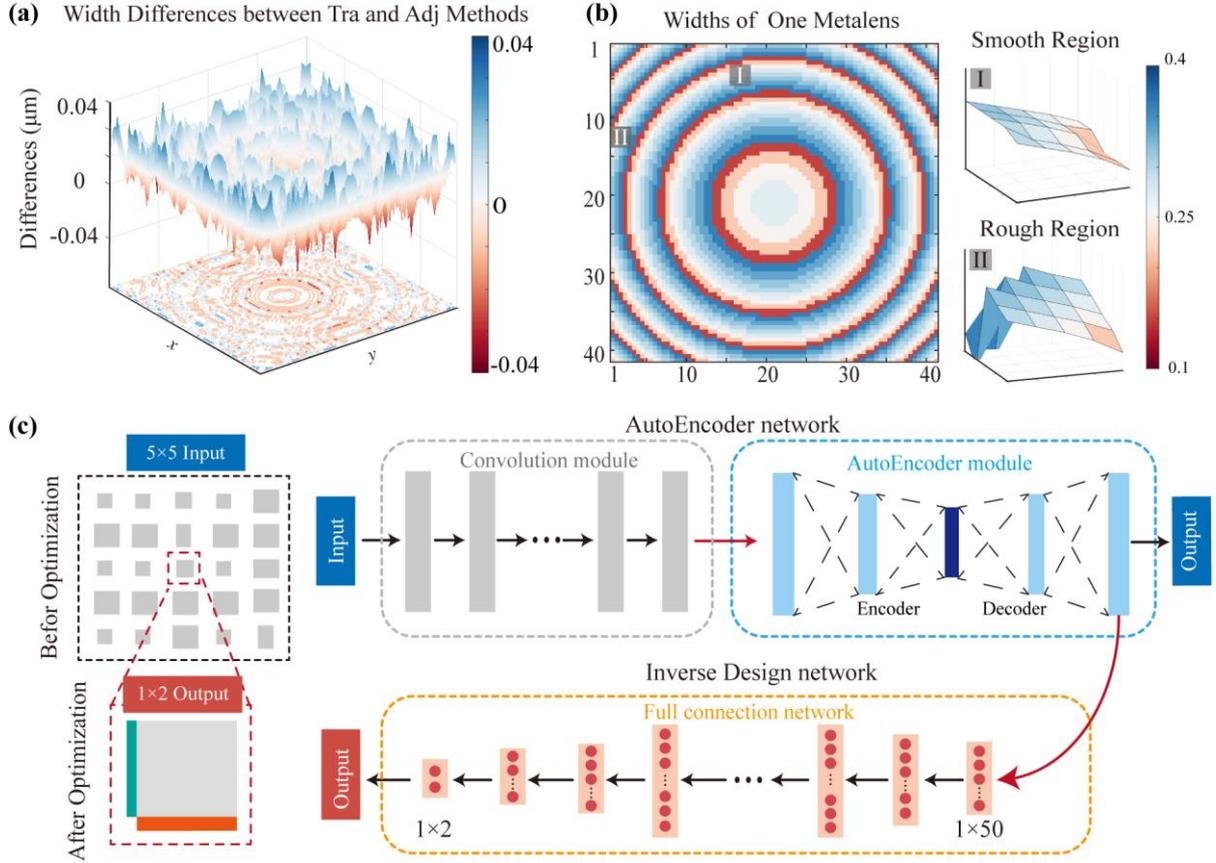

**Figure 2** Design method for the large-scale metalens. **a** The differences in widths distribution for the meta-atoms before and after optimization. **b** The widths distribution of metalens, with the smooth region and rough region corresponding to translucent boxs I and II, respectively. **c** The framework of the optimized network where the A-network expands the information space of sampled data, and the weak coupling strength structures are filtered by the I-network.

$$\varphi = \frac{2\pi}{\lambda}(f - \sqrt{f^2 + (x-x_0)^2 + (y-y_0)^2}) \qquad (1)$$

where $f$ is the focal length, $\lambda$ is the working wavelength, and $x$ and $y$ are the spatial positions on the lens. $x_0$ and $y_0$ correspond to the displacement of the focal spot from the center of the metalens. There is a one-to-one mapping between each phase and the size of the meta-atom, and the phase accuracy depends on the kinds of meta-atoms. At last, the Adj method is used to optimize the real electric field distribution. Although the Adj method is also useful for random distribution structures, it will enforce the one-to-one mapping to become many-to-one, bringing greater challenges for the network training. Taking the metalens with a diameter of 40.5 μm (81×81 meta-atoms) as an example, after 20 optimization iterations, all the meta-atoms sizes have been effectively corrected. As shown in Fig.2(a), there is a difference of ±40 nm in widths between initial and optimized meta-atoms. Finally, metalenses have Strehl ratios of 0.96/0.958 and 0.95/0.945 for $x$/$y$-polarized light. The figure of merit (FoM) is desired transmitted field distribution which also indicates the Strehl ratio of the metalens.

Next, metalenses before and after optimization is used to build data sets, that is, to carry out data pre-processing steps which can be separated into 2 steps. Firstly, the initial metalenses are divided into many super meta-atoms whose widths/lengths ultimately become the input data sets. Considering that a coupling effect occurs within the distance range between a meta-atom and its surrounding two layers of meta-atoms, the size of a super meta-atom is 5×5 with a sliced interval of 1. This means the design-loop will work on a 5×5 array with overlap. Since the initial metalens is isotropic, the input data sets can only consist of widths or lengths. Secondly, the output data sets are derived from the optimized metalens. One output data set is 1×2 vector data consisting of the width and length of one meta-atom, which is the modified meta-atom positioned at the center of the super meta-atoms. Because the FoM of Adj method is based on both *x*- and *y*- polarized incident light, it makes the meta-atom eventually anisotropic, leading to the different width and length. The initial widths/lengths distribution of a metalens is shown in Figure 2(b). There are two types of super meta-atoms, namely continuous change (smooth) box I and discontinuous change (rough) box II. The features of a smooth region are that the widths/lengths of the meta-atoms in the same region are increased, decreased, or remain unchanged. On the contrary, regions, where the widths/lengths change do not follow the above three patterns are referred to as rough regions. As a result, the metalens is divided into many super meta-atoms. Finally, we have 9178 data sets consist of two metalenses, of which 3960 data sets belong to smooth regions and 5218 belong to rough region. In the following, the validation loss of the network will prove that thousands of data sets are enough for our network. Figure 2(c) shows the framework of the optimized network, which consists of an AutoEncoder network (A-network) and an inverse design network (I-network). The input data sets are encoded and decoded, with the output data of the A-network being equal to the input data, promoting information depth mining. The latent layer carries all the information of the original input data in a different form, and the outputs of the A-network are coupled to the I-network which is utilized to generate the optimized meta-atoms.

## 3 Results and Discussion

Two networks were trained for dealing with smooth and rough regions of data sets. The architectures of A-network and I-network are shown in Supplementary Information (see Section S2). The mean-absolute-error (MAE) loss function is used to measure the difference between the predicted values and the real values, and helps the optimizer adjust the model parameters to improve the prediction accuracy. After 5000 iterations, the MAEs for smooth and rough region networks are $3\times10^{-3}$ and $2\times10^{-3}$ (see Supplementary Information, Section S3), corresponding to output errors of 3 nm and 2 nm (both within the fabrication error), respectively. The above networks have the advantage of directly obtaining the output from the fully connected neural (FCN) network. The loss of the FCN is $7\times10^{-3}$ which is greater than the loss of our network. The details of the FCN network and its performance are shown in Supplementary Information (see Section S4).

The same as the design of the initial small-scale metalens, the large-scale metalens could also be designed by the same 8 kinds of meta-atoms. To improve the focusing efficiency of large-scale metalens, it will be sliced into many super meta-atoms each consisting of 5×5 meta-atoms, and the input data to the network are widths/lengths of a super meta-atom. Using the trained DL network, we get many new output data sets, each of which consists of the lengths and widths of the new meta-atom. Finally, the output data sets can be arranged on the same path as the input slice path to form a new metalens. To verify the performance of the metalens optimized by the Adj method and our method, we optimized the same linear polarization-multiplexed metalens with a diameter of 50.5 µm using the aforementioned methods. The electric field distribution of the designed metalenses can be obtained through FDTD Solution software and the vector diffraction theory. The full-wave simulation results proved that the focusing efficiencies of both metalenses mentioned above are higher than the initial metalens at the working wavelength of 1550 nm.

Figures 3(a)-(b) shows electric field distributions and the intensity profiles for different metalens on the $xy$ focal plane and $xz$ plane. After 15 days of 20 optimization iterations, the relative and absolute focusing efficiencies of the metalens have been improved to 87%/86.8% and 78%/77.9% with the Adj method for $x$- and $y$-polarization, respectively. The final relative and absolute focusing efficiencies of the metalens designed with the proposed method are 86.5%/87.1% and 76%/76.6% for $x$- and $y$-polarized light, and the optimization time is within 17 s. Here, the absolute focusing efficiency is defined as the ratio between the light intensity from the focal spot and the incident intensity, which can also be calculated by the product of transmission efficiency and the relative efficiency of the device. The relative focusing efficiency is defined as the ratio between the light intensity from the focal spot and the focal plane. The Strehl ratios of $x$- and $y$-polarized light are 0.95/0.957 and 0.96/0.96 for the metalenses designed by the DL method and Adj method, respectively. Figures 3(c)-(d) show the widths distribution on the $y=0$ plane ($x>0$) designed by three method and compare them with the initial structure distribution on the $y=0$ plane ($x>0$). As shown in Fig. 3(d), there is a difference between the width obtained by the above two methods and the width obtained by the traditional method. Although the width changes in each pixel are inconsistent, the metalenses formed by them all exhibit high focusing efficiencies, which may result from non-convex optimization. The above simulation results have proved that the method base on the DL network

can achieve vector optimization for large-scale metalens while significantly reducing the required time by five orders of magnitude.

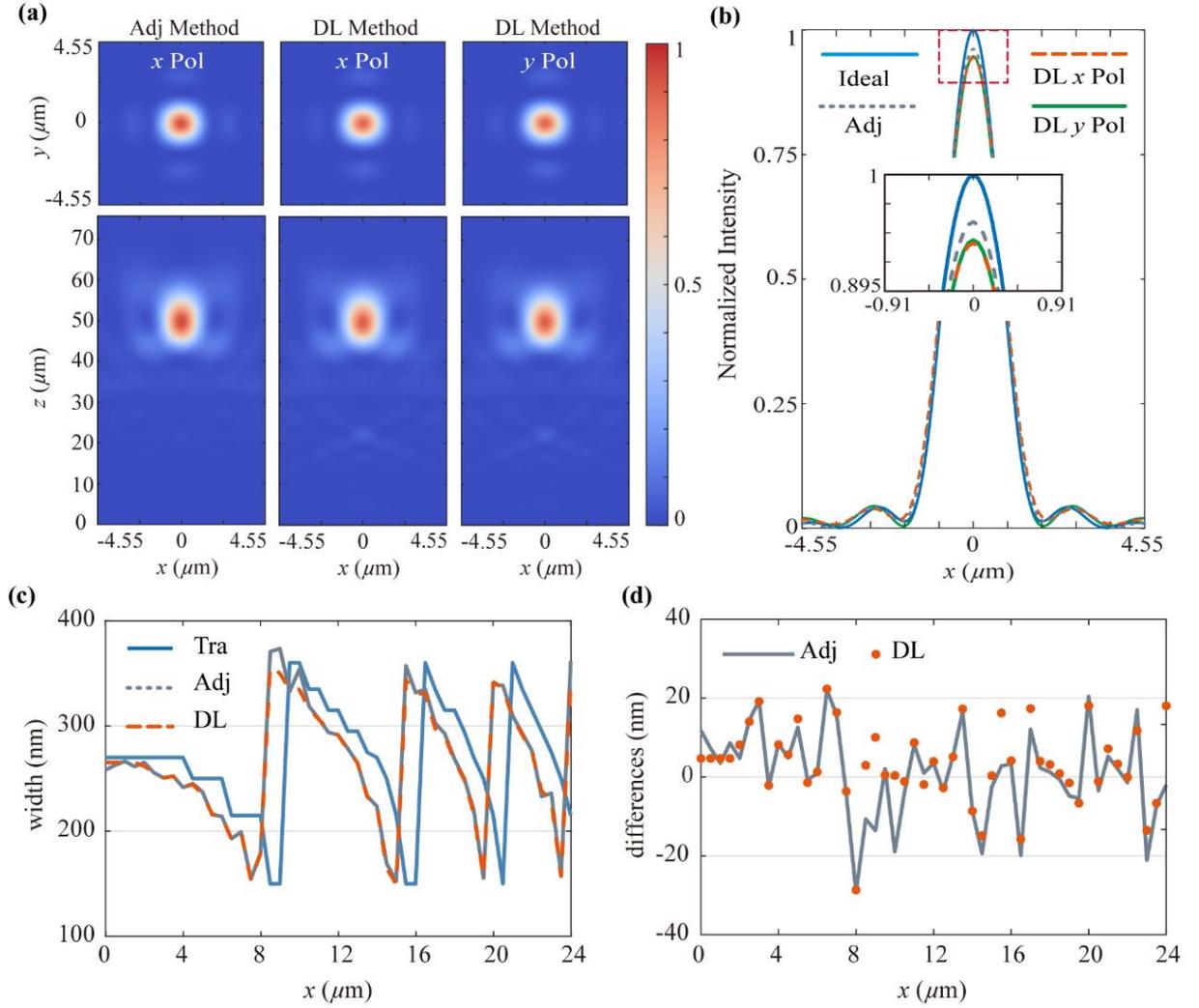

**Figure 3** Simulation results of Adj and DL methods. **a** Electric field distributions in the *xy* and *xz* planes designed by DL method for *x*- and *y*-polarized light, and the Adj method for *x*-polarized, respectively. **b** Electric intensity profiles of the focal spot designed by theory, Adj, and DL methods, respectively. **c** Width distributions of metalenses on *y*=0 plane (*x*>0) designed by the traditional (Tra), Adj, and the proposed methods, respectively. **d** Width differences between the initial metalens and the optimized metalenses designed by the Adj method and our methods.

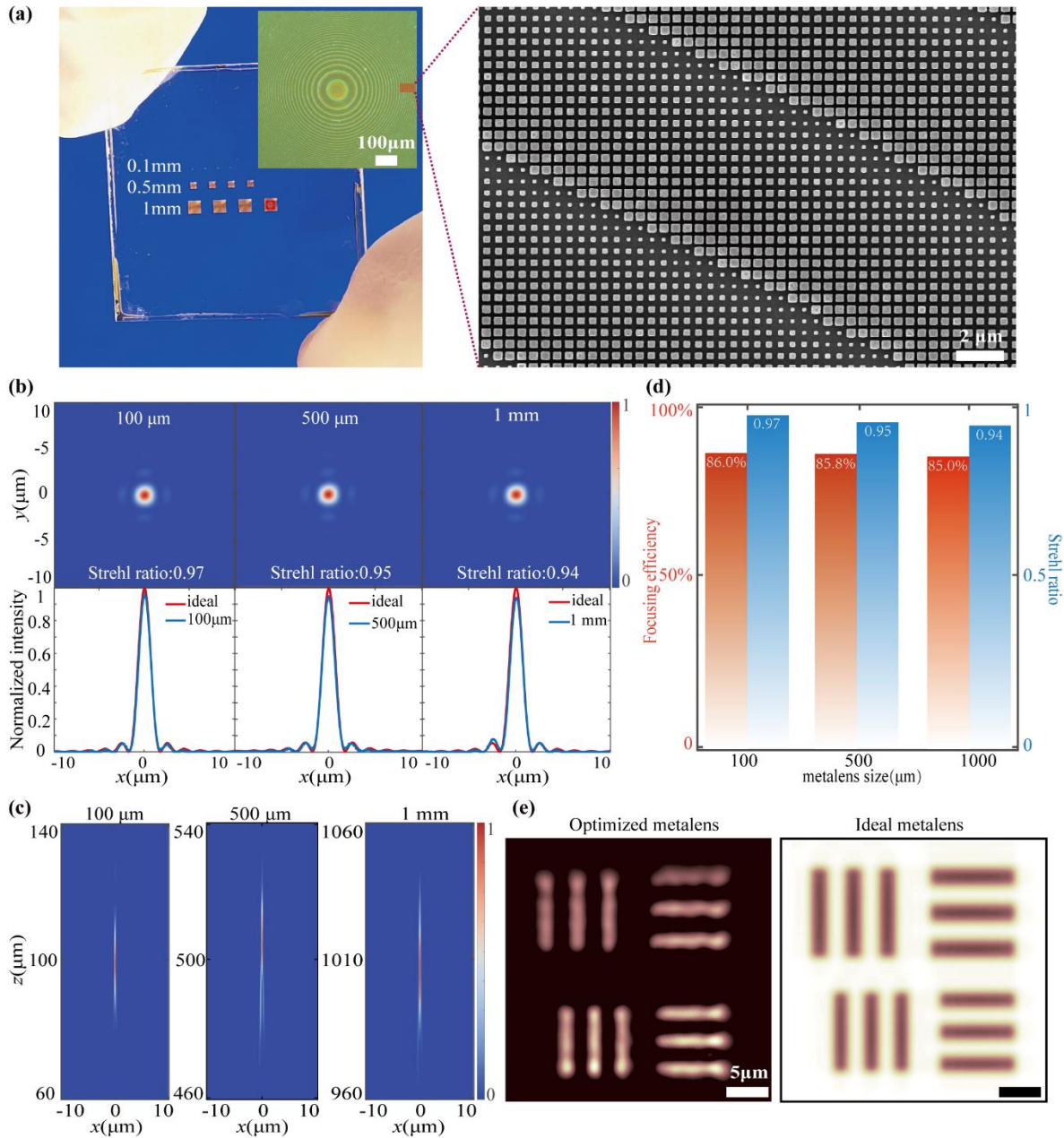

**Figure 4.** Experimental results of optimized metalenses. **a** The left image shows an overview of the device, with each size corresponding to four exposure doses. The inset is the optical microscope image, with a scale bar of 100 µm. The right image is the scanning electron microscope image, with a scale bar of 2 µm. **b** The first row shows the focal plane intensity distributions of the three metalenses (100.5 µm, 500.5 µm, and 1.0005 mm, from left to right, respectively). The second row shows the normalized focal intensity along the *x*-axis at the focal plane of the three metalenses. **c** Focal intensity distributions in the *xz* plane at the three metalenses. **d** Relative focusing efficiencies and Strehl ratios of three metalenses. **e** Imaging results of elements #5 and #6 from group #7 of the USAF resolution target at 1 mm size optimized metalens (left) and 1mm size ideal metalens (right). The scale bar of the two figures is 5 µm.

To verify the optimization capability of the network for larger-scale metalenses, two metalenses with diameters of 60.5 µm and 100.5 µm were optimized, respectively. The electric field distributions are shown in Supplementary Information section 4 (see Figure S5), and the metalenses have the efficiencies and Strehl ratios of 86.6%, 86.2%, 0.97, and 0.96, respectively. Furthermore, to validate the generalization ability of the network, three metalenses with NAs of 0.3, 0.35, and 0.4 were optimized, and the corresponding electric field distributions are shown in Supplementary Information Section 5 (see Figure S6). The above metalenses have efficiencies and Strehl ratios of 86.4%, 86.6%, 87.7%, 0.94, 0.95, and 0.95, respectively. The above results have demonstrated that the trained network can optimize large-scale metasurfaces with F-numbers ≥ 1 when the data sets consists of metalens with an F-number of 1.

For further proof of concept, the optimized metalenses with diameters of 100.5 µm, 500.5 µm, and 1.0005 mm with F-numbers of 1 were fabricated using the electron beam lithography technology on the silicon-on-sapphire substrate. Figure 4a shows the optical microscope and the scanning electron microscope (SEM) images of the fabricated metalenses, where the relative efficiencies of the metalenses under the illumination of the linear polarized laser beam are measured. Considering the photoelectric response of the CCD to the incident laser energy shows certain nonlinear response characteristics [56], a power meter is used to detect the relative efficiency (see Figure S9, Supplementary Information for experiment set-up). A focusing lens and a precision pinhole are used to constrain the size of the incident light and focused spot, respectively. Lens modulation ensures that the beam size of the incident light is smaller than that of the metalens. Additionally, a precision pinhole is placed behind the metalens to filter out power outside the focus range.

The fabricated metalenses all have an NA of 0.44, and the ideal FWHMs of the metalenses are ~1.57 µm, calculated by the vectorial angle spectrum method. The diameter of the precision pinhole is 10±1 µm, matching the size of the effective focal spot (a circular region centered on the peak intensity with a radius of treble FWHMs, and the diameter is 9.42 µm). The power meter is placed behind the precision pinhole to measure the focusing power and the focal plane power will be measured when the precision pinhole is removed. Considering the short focal length of a metalens with a diameter of 100.5 μm, an objective lens with 100 × magnification is used to pair with a tube lens to magnify the image of the focal spot. An aperture with diameter of 1 mm is chosen to replace the precision pinhole, and other processes are the same as above. Under the illumination of a 1550 nm laser, the relative efficiency of the DL-optimized metalens with a diameter of 1 mm is 85%, which is 93.4% of the ideal relative focusing efficiency.

Furthermore, the distribution of the focal spot on *xy* and *xz* planes is shown in Figures 4b and c. Figure 4b shows the normalized focusing intensity along the optical axis of the ideal metalens and the DL-optimized metalens, where the sub-maximum intensity of the ideal focal spot is used to normalize the intensity of the DL-optimized focal spot and the ideal focal spot. The Strehl ratio is defined as the ratio of the maximum intensity of the DL-optimized focal spot and the ideal focal spot. Objectives with large NA=0.7 are used to avoid the limited field of view of transmitted light. The experimental set-up is shown in Figure S11 (Supplementary information). The imaging performance of the metalens (with a diameter of ~1 mm) is further characterized by the USAF 1951 resolution test chart (see Figure S12, Supplementary Information for experiment set-up), as shown in Figure 4(e). The imaging result is obtained from group 7, where the smallest feature size is 2.19 μm of element No.6, and it can be clearly resolved. As a comparison, the imaging result of the ideal metalens (with a diameter of ~1 mm) is demonstrated, which is obtained through the convolution of ideal metalens focusing spot and resolution test chart. Experimental results show that the electromagnetic properties of metalenses designed by our DL method are close to those of ideal metalenses.

**4 Conclusions**

In this work, an "intelligent optimizer" is proposed to address the difficulty of quickly optimizing large-scale 2D metalens. The numerical and experimental results demonstrate the promising potential of the "intelligence optimizer" in developing high-efficiency metalens technology. We successfully optimized the geometry parameters of a high-performance metalens with a diameter of mm-scale using the geometry parameter information of a high-performance metalens with a diameter of μm-scale. The optimized metalens, with diameter of 1 mm consist of $4\times10^6$ meta-atoms, can be optimized only in 11 minutes and 20 seconds. It has a relative focusing efficiency of 93.4% (as compared to ideal focusing) and a Strehl ratio of 0.94. In addition, as aforementioned, the optimizer has the ability to scale downwards. Each size of meta-atoms in the metalens represents a phase, which means that optimizing the size of meta-atoms is also an optimization of the phase. The proposed optimization method is not limited to polarization-multiplexed metalenses, and it can also be used for other meta-devices, such as achromatic metalenses, wide-angle metalenses, orbital angular momentum metalenses, etc.

## 5 Methods

The full-wave simulations of the metalenses were performed with FDTD Solutions. Both $x$- and $y$-polarized plane waves were normally incident from the bottom of the metasurface in the $+z$ direction, with the boundary conditions set as a perfectly matched layer (PML) along $x$, $y$, and $z$ directions. The working wavelength was 1550 nm, and the mesh size was 50 nm. The data sets were established using the conventional FDTD solver, and it took ~10 days to complete 20 iterations of optimization. The DL network was trained using Tensorflow and the training time is about 30~40 minutes. The optimization time for metalenses with diameters of 100.5 μm, 500.5 μm, and 1000.5 μm are the 20 seconds, 2 minutes 45 seconds, and 11 minutes 20 seconds, respectively. Except from the simulation of the electric field, all the computation is dealt with on the Pycharm software with an Intel Xeon Gold 6254 CPU Processor with a base clock of 3.09 GHz, 128 GB of RAM. The metasurface was fabricated by electron beam lithography techniques. The fabrication flow chart and the SEM images of the fabricated metalens with a diameter of 1 mm are shown in Supplementary S7 and S8. The electric field and the imaging results of the metalens were characterized by the setup (see Supplementary S9, S11, and S12 for detailed description).

**Abbreviations**
Tra: Traditional method; Adj: Adjoint optimization; DL: Deep learning; FoM: Figure of merit; FDTD: Finite difference time domain; SEM: Scanning electron microscope

**Ethics approval and consent to participate**
Not applicable.

**Consent for publication**
Not applicable.

**Availability of data and materials**
All data generated or analysed during this study are included in this published article and its additional files.

**Competing interests**
The authors declare that they have no competing interests.

**Funding**
This work was supported by the National Natural Science Foundation of China (No. 61975210, 62175242, 62105338, and 62222513), Sichuan Science and Technology Program (2020YFJ0001), the Postdoctoral Science Foundation of Sichuan (J22S001).

**Authors' contributions**
MB. P., YL. H. and Y. L. proposed the idea, YL. H. and Y. L. carried out the simulations and prepared the manuscript. Q. H. fabricated the metasurfaces. Y. L., F. Z., JJ. J. measured the optical properties of the metasurfaces. MB. P., XG. L. supervised the project. All the authors analyzed the data and discussed the results. The authors read and approved the final manuscript.

**Acknowledgements**
Not applicable.

**Authors' information**
[1]State Key Laboratory of Optical Technologies on Nano-Fabrication and Micro-Engineering, Institute of Optics and Electronics, Chinese Academy of Sciences, Chengdu 610209, China.
[2]Research center on vector optical fields, Institute of Optics and Electronics, Chinese Academy of Sciences, Chengdu 610209, China. [3]School of Optoelectronics, University of Chinese Academy of Sciences, Beijing 100049, China. [4]Tianfu Xinglong Lake Laboratory, Chengdu 610299, China.

# Supplementary Information

**Physics-data-driven intelligent optimization for large-scale meta-devices**

Yingli Ha[1,2,†], Yu Luo[1,2,†], Mingbo Pu[1,2,3,*], Fei Zhang[1,2], Qiong He[1], Jinjin Jin[1], Mingfeng Xu[1,2,3], Xiaogang Li[4], Yinghui Guo[1,2,3], Xiong Li[1,3], Xiaoliang Ma[1,3], and Xiangang Luo[1,3*]

[1] State Key Laboratory of Optical Technologies on Nano-Fabrication and Micro-Engineering, Institute of Optics and Electronics, Chinese Academy of Sciences, Chengdu 610209, China.

[2] Research Center on Vector Optical Fields, Institute of Optics and Electronics, Chinese Academy of Sciences, Chengdu 610209, China.

[3] School of Optoelectronics, University of Chinese Academy of Sciences, Beijing 100049, China.

[4] Tianfu Xinglong Lake Laboratory, Chengdu 610299, China.

[†] These authors contributed equally: Yingli Ha, Yu Luo.

[*] Correspondence: pmb@ioe.ac.cn ; lxg@ioe.ac.cn

**Section 1. Principle of adjoint-based shape optimization**

The adjoint optimization method is a kind of gradient descent algorithm based on adjoint field physics. In each optimization process, the gradient information of the objective function in the whole parameter optimization space can be obtained only by calculating the forward field and the adjoint field once.

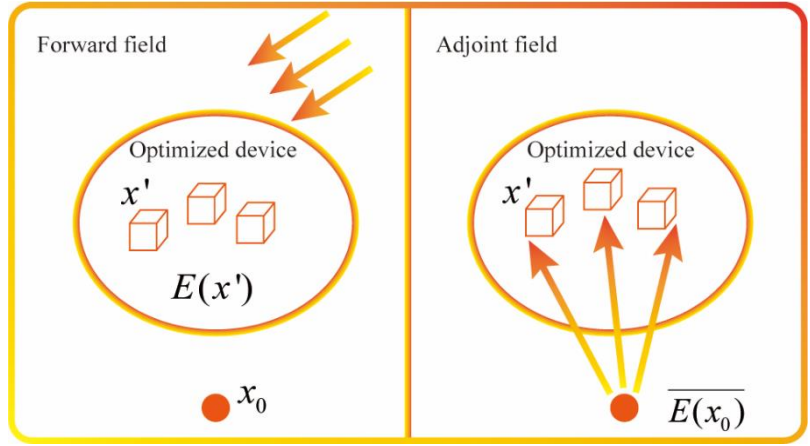

**Figure S1** Schematic diagram of adjoint optimization.

Adjoint-based shape optimization is to optimize the geometric dimensions of binary patterns. As shown in Figure S2, the initial structure boundary is a solid line, and the updated boundary is a dashed line. For a given deformation, the shadow area is $\psi$ express. This figure can be regarded as a two-dimensional section of the deformation process of three-dimensional graphics. To region $\psi$, the integration needs to be carried out along the boundary of the initial pattern, and the deformation of the pattern needs to be along the normal direction of the boundary.

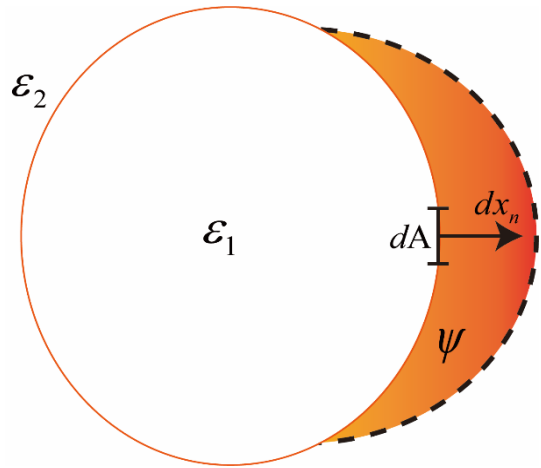

**Figure S2** Schematic diagram of adjoint-based shape optimization.

The figure of merit (FOM) can be written as:

$$\delta FOM = 2\operatorname{Re} \int_\psi \mathbf{P}^{ind}(x') \cdot \mathbf{E}^A(x') d^3 x' \qquad (1)$$

where $\mathbf{P}^{ind}$ is the induced dipole and $\mathbf{E}^{A}(x')$ is the adjoint electric field at the object location $x'$. According to the figure S2, $dA$ is the derivative along the surface, and $dx_n$ is the normal derivative, equation (1) can be written as:

$$\delta FOM = 2\text{Re}\int dA \int dx_n [\mathbf{P}^{ind}(x') \cdot \mathbf{E}^{a}(x')] \tag{2}$$

When the deformation tends to zero:

$$\int x_n \to \delta x_n(x') \tag{3}$$

where denotes the shape variable of the normal direction of each point on the boundary. So:

$$\delta FOM = 2\text{Re}\int \delta x_n(x')\mathbf{P}^{ind}(x') \cdot \mathbf{E}^{a}(x')dS \tag{4}$$

According to the boundary conditions of Maxwell's equations, the tangential direction of $\mathbf{E}$ is continuous and the normal direction of $\mathbf{D}$ is continuous. Only in the continuous field, Equation (4) is meaningful. So the induced dipole can be written as:

$$\mathbf{P}^{ind}(x') = (\varepsilon_2 - \varepsilon_1)\mathbf{E}^{new}(x') \tag{5}$$

where $\mathbf{E}^{new}$ is the steady-state electric field of the deformed shape. Because even a small change in the boundary, the discontinuous component of $\mathbf{E}$ will change dramatically, so $\mathbf{E}^{new} \cong \mathbf{E}^{old}$ is not true, so:

$$\begin{aligned}\mathbf{E}^{new} &= \mathbf{E}^{old}_{\parallel} + \delta\mathbf{E}_{\parallel} + \frac{\mathbf{D}^{old}_{\perp} + \delta\mathbf{D}_{\perp}}{\varepsilon_2} \\ &\cong \mathbf{E}^{old}_{\parallel} + \frac{\mathbf{D}^{old}_{\perp}}{\varepsilon_2}\end{aligned} \tag{6}$$

Similarly, the adjoint electric field can be written as follows:

$$\mathbf{E}^{a} = \mathbf{E}^{a}_{\parallel} + \frac{\mathbf{D}^{a}_{\perp}}{\varepsilon_1} \tag{7}$$

So the change of FOM is written as:

$$\delta FOM = 2\text{Re}\int \delta x_n(x')[(\varepsilon_2 - \varepsilon_1)\mathbf{E}_\parallel^f(x')\cdot \mathbf{E}_\parallel^a(x') + (\frac{1}{\varepsilon_1} - \frac{1}{\varepsilon_2})\mathbf{D}_\perp^f(x')\cdot \mathbf{D}_\perp^a(x')]dS \qquad (8)$$

From calculus:

$$\delta FOM = \sum_i \frac{\partial FOM}{\partial x_i}\delta x_i \qquad (9)$$

For an optimization algorithm based on gradient:

$$\delta x_i = \frac{\partial FOM}{\partial x_i} \qquad (10)$$

So, for all points $x'$ prime on the boundary, the deformation in the normal direction can be expressed as:

$$\delta x_n(x') = 2\text{Re}[(\varepsilon_2 - \varepsilon_1)\mathbf{E}_\parallel^f(x')\cdot \mathbf{E}_\parallel^a(x') + (\frac{1}{\varepsilon_1} - \frac{1}{\varepsilon_2})\mathbf{D}_\perp^f(x')\cdot \mathbf{D}_\perp^a(x')] \qquad (11)$$

$\mathbf{E}_\parallel^f$ and $\mathbf{E}_\parallel^a$ denote the tangential components of the electric field obtained in the forward and adjoint simulations, respectively. $\mathbf{D}_\perp^f$ and $\mathbf{D}_\perp^a$ denote the normal components of the potential shift-vector in the forward and adjoint simulations, respectively.

**Section 2. Inverse Design Network Structure Distribution.**

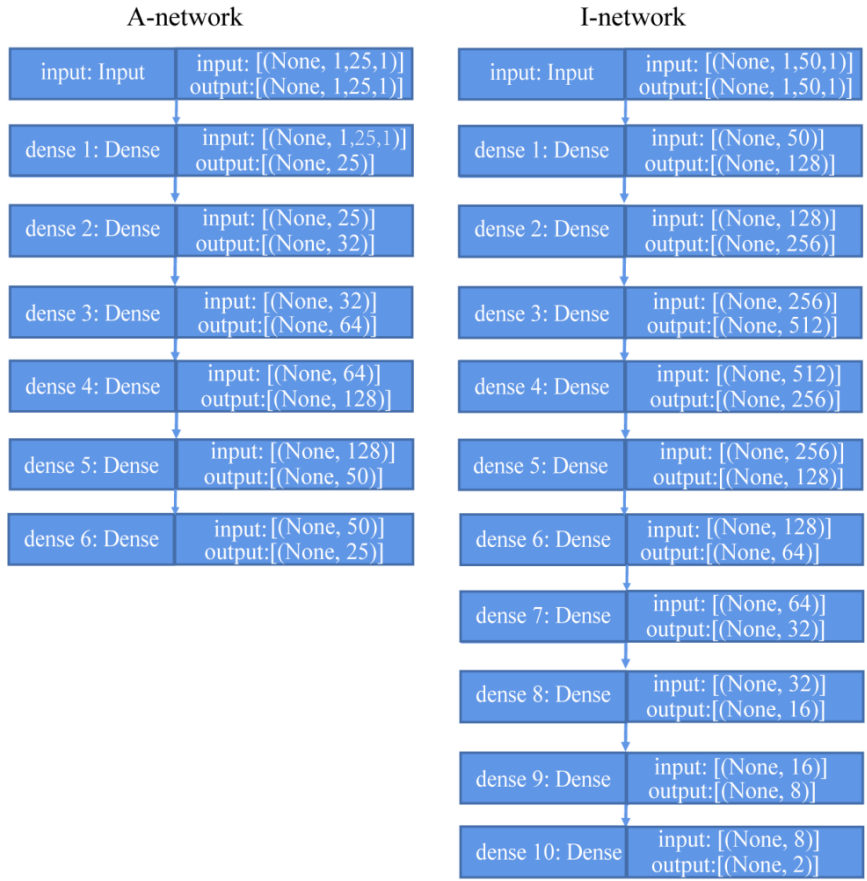

**Figure S3** Network architectures. (a) A-network; (b) I-network.

**Section 3. Training loss for smooth and rough regions.**

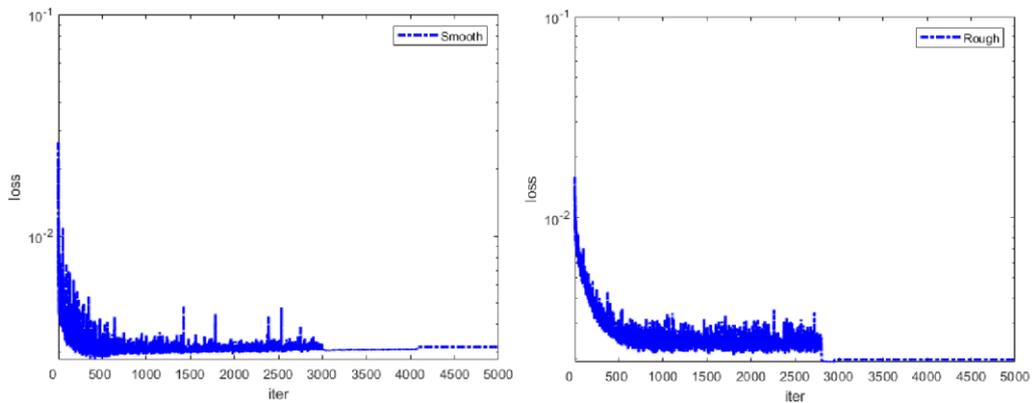

**Figure S4** (a)Loss of different iterations for smooth region; (b) Loss of different iterations for rough region.

**Section 4. Overview of the FCN network architecture and training procedure.**

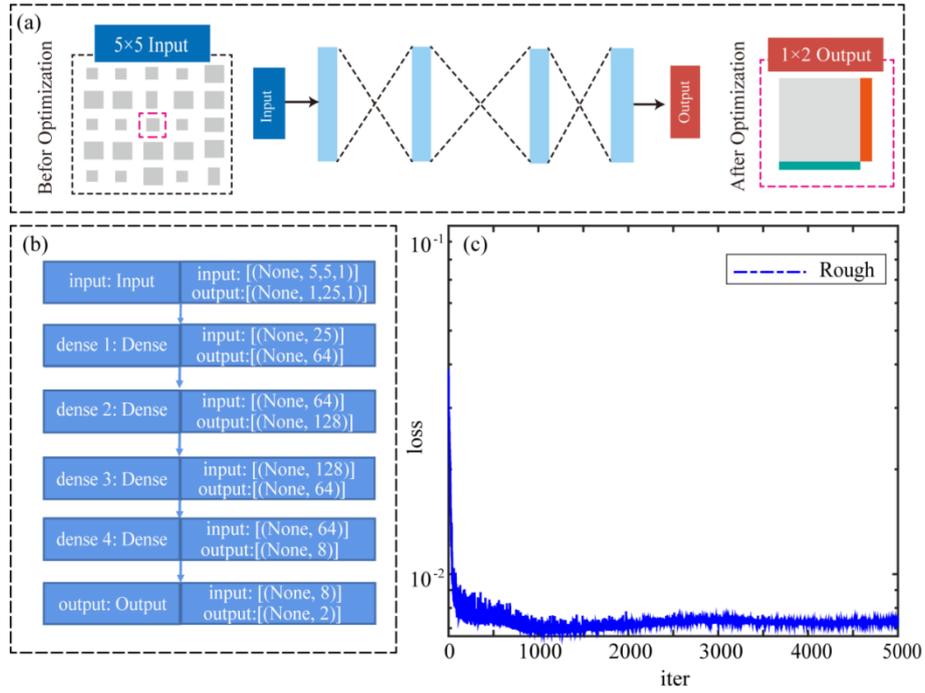

**Figure S5.** Overview of the FCN network architecture and training procedure. (a) Schematic of the FCN network. (b) The network architectures of the FCN. (c) The MAE loss for the smooth region.

The fully connected neural (FCN) network is shown in Fig. S5 (a), and one input data with a size of 5×5 is composed of the width/length of a super meta-atoms, while the output data with a size of 1 × 2 is composed of the width and length of the center meta-atom. As shown in Fig. S5 (b), the network architectures of the FCN contain 5 layers, and the filter size of each layer is shown in Fig. S5 (b). Figure S5 (c) depicts the simulated dependence of loss value on the iteration number, and the loss eventually reaches the minimum (0.007) at the iteration of 5000.

**Section 5. Simulated results of metalens with different diameters.**

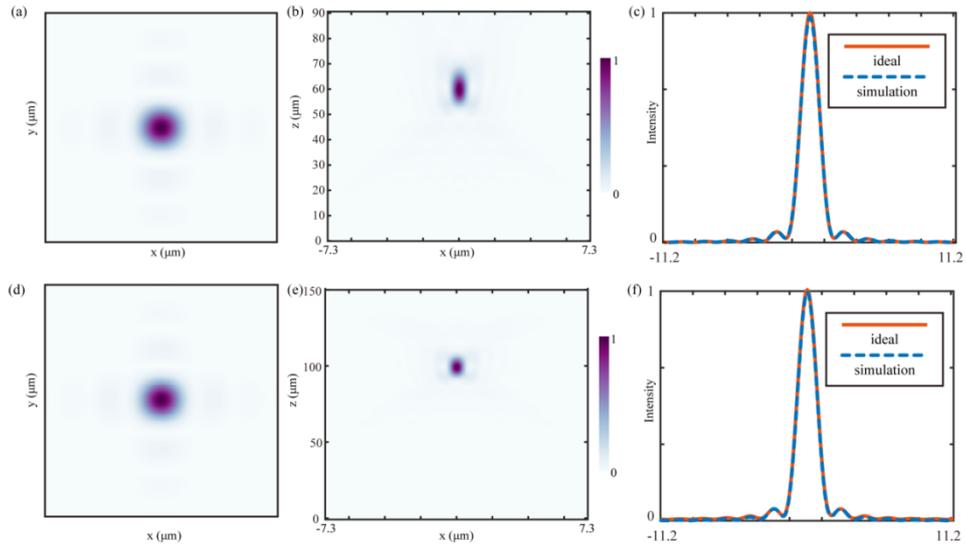

**Figure S5** (a)-(c) Electric field distributions on *xy* and *xz* plane, and electric intensities of the focal spot for the metalens with the diameter of 60 μm; (d)-(f) Electric field distributions on *xy* and *xz* plane, and electric intensities of the focal spot for the metalens with the diameter of 100 μm.

**Section 6. Simulated results of metalens with different NA.**

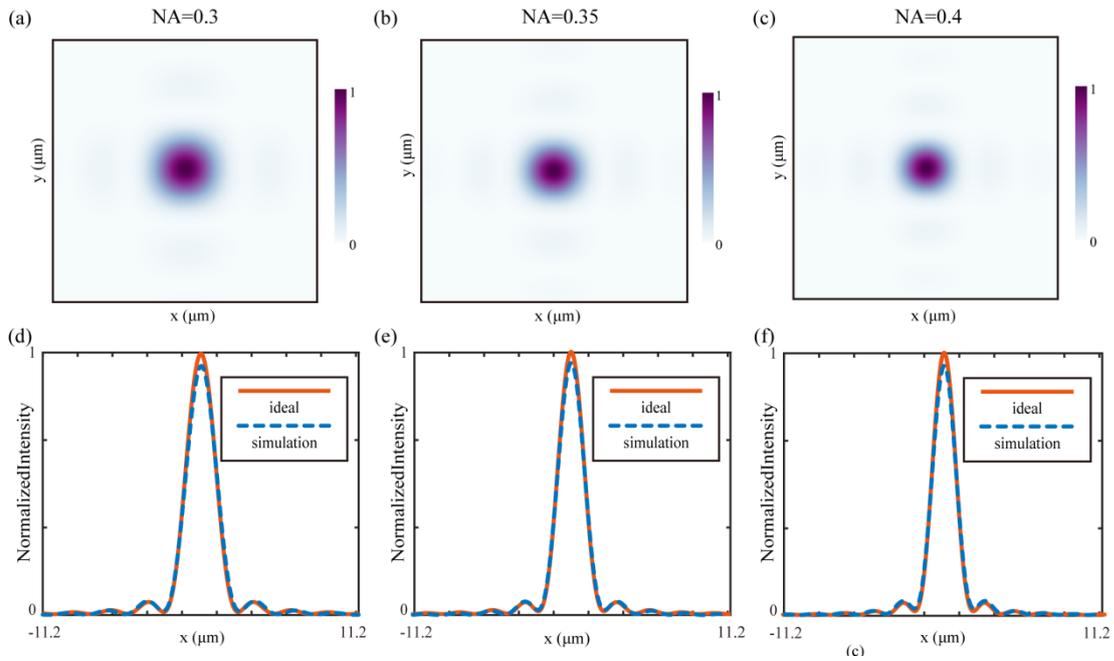

**Figure S6** (a)-(c) Electric field distributions on *xy* plane for the metalens with the NA of 0.3, 0.35, and 0.4; (d)-(f) Normalized electric intensities of the focal spot for the metalens with the

NA of 0.3, 0.35, and 0.4.

**Section 7. The metalens fabrication process**

The schematic of the metalens fabrication process is presented in Figure S7. First, a 500 μm-thick Silicon on Sapphire (SOS) substrate was cleaned by the piranha solution followed by ultrasonic cleaning under acetone and isopropanol respectively. Next, a 100 nm-thick negative electron resist (ma-N2401, Micro resist technology) was spin-coated on the substrate and baked at 90 °C for 60 s. The electron beam patterning was performed by Elionix ELS-F125. The exposed sample was developed at room temperature for 10 s (ma-D 525, Micro resist technology). Then the inductively coupled plasma reactive ion etching (SENTECH SI 500) was employed to etch the silicon structures with SF6 and C4F8 mixture and to remove the leaving electron resist.

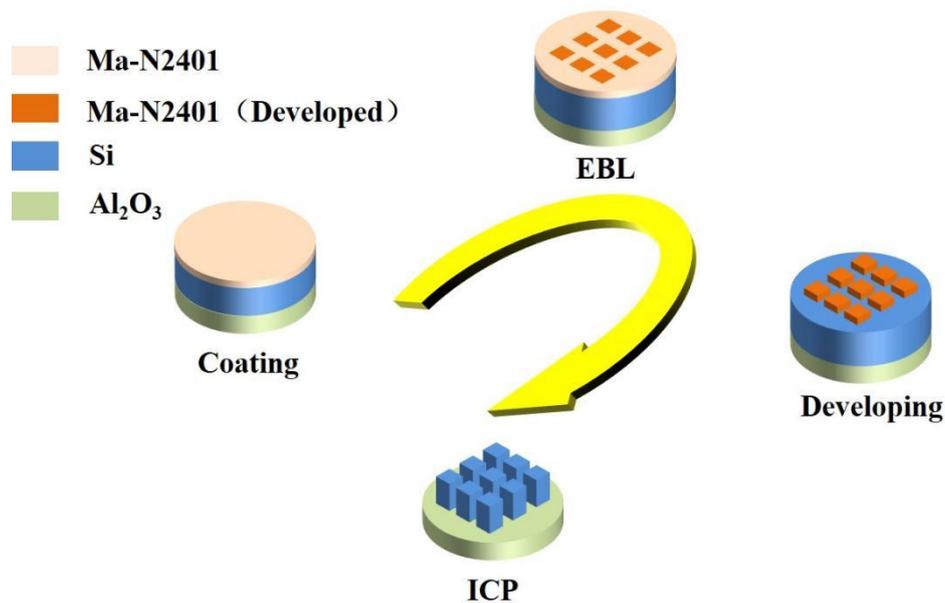

**Figure S7** Schematic of the metalens fabrication process.

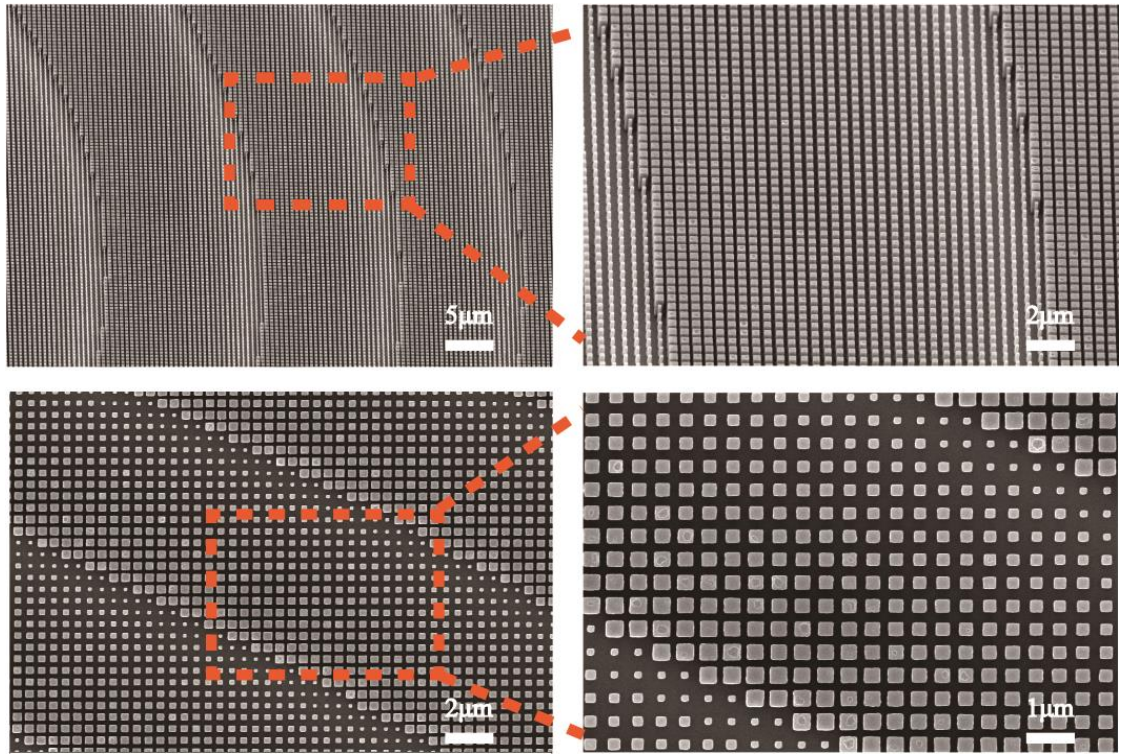

**Figure S8** Scanning electron microscope image of the metalens with diameter of 1 mm.

## Section 8. Optical setup for metalens measurements

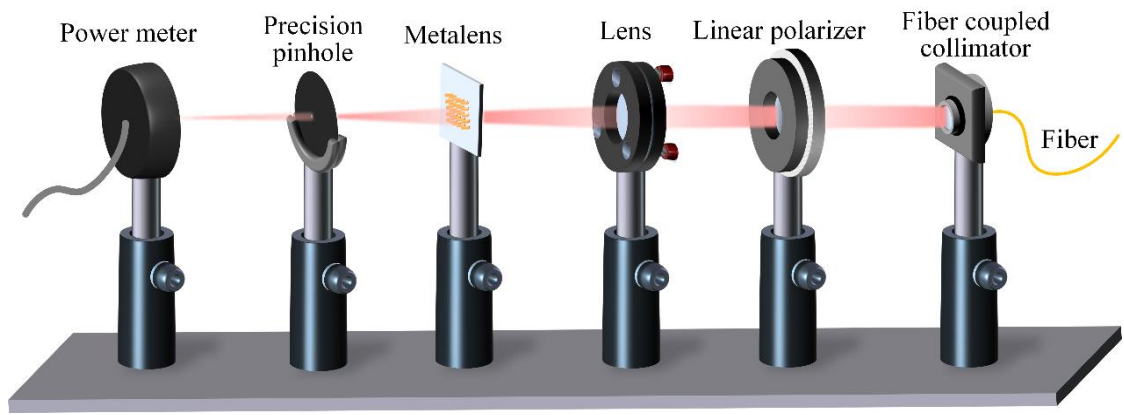

**Figure S9** Schematic of the optical setup used to measure the focusing efficiency of the metalens. The fiber laser output 1550nm line polarization beam is collimated by a fiber coupled collimator (Thorlabs, PAF2-A7C), liner polarizer is used to adjust the polarization state of the beam and adjust the laser intensity. A lens is used to adjust the size of laser beam. The adjusted laser beam with size same as the diameter of metalens. A precision pinhole (Edmund,

Unmounted) is placed at metalens's focal length, the precision pinhole with 10μm diameter to filter focus energy. A power meter is used to catch the energy incident on it.

We have measured relative focusing efficiency of the three sizes of metalens with different exposure doses. The relative focusing efficiency is defined as the ratio of the focusing power to the power of the focal plane, the focusing range is a circular area centered on the peak intensity, with a radius of treble full-width-half-maximums (FWHMs).

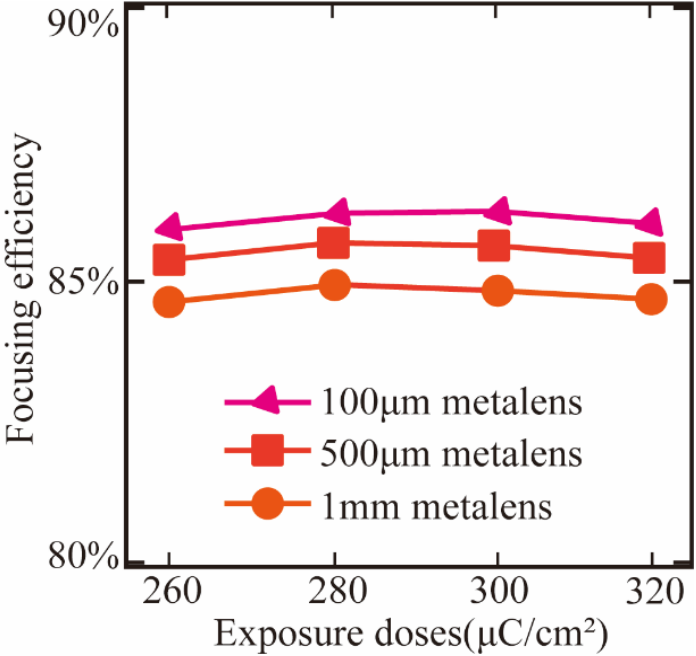

**Figure S10** Measured focusing efficiency of three size metalens with different exposure doses.

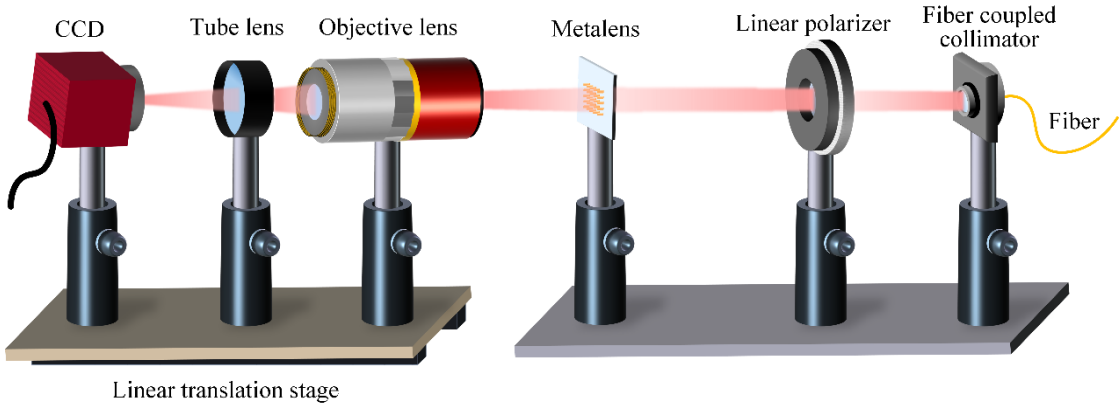

**Figure S11** Schematic of the optical setup used to measure the focusing spot of the metalens. The fiber laser output 1550nm line polarization beam is collimated by a fiber coupled collimator (Thorlabs, PAF2-A7C), liner polarizer is used to adjust the polarization state of the beam, then incident on the metalens. The microscopic imaging system is composed of a objective lens (Mitutoyo, 100× magnification), a tube lens (Edmund, MT-40) and a CCD camera (Allied Visio, G130), used to characterization the three-dimensional intensity distribution of the focal spot. The microscopic imaging system is installed on a linear translation stage.

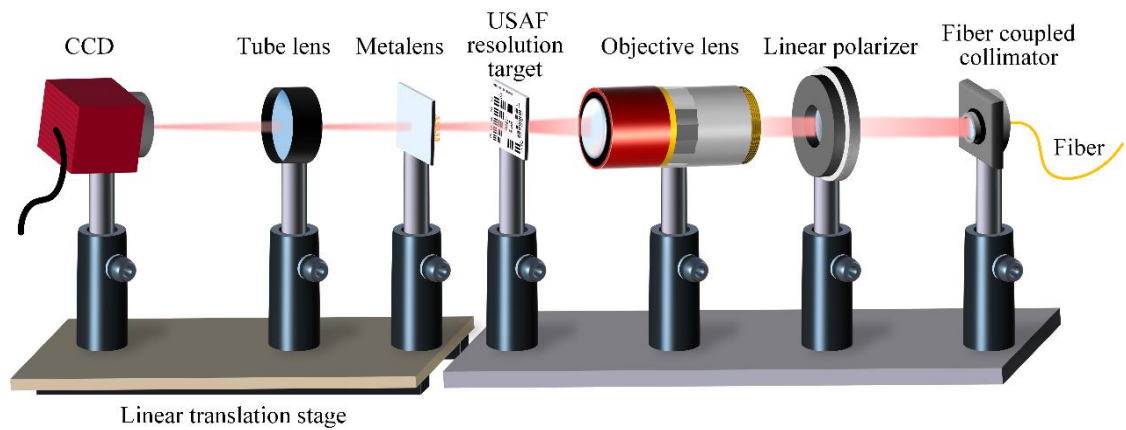

**Figure S12** Schematic of the optical setup used to characterize the imaging performance of the metalens. The fiber laser output 1550nm line polarization beam is collimated by a fiber coupled collimator (Thorlabs, PAF2-A7C), liner polarizer is used to adjust the polarization state of the beam, then incident on an objective lens (Mitutoyo objective, 10× magnification). The laser is focused by the objective lens and illuminated the United States Air Force resolution target(USAF resolution target). The metalens is placed parallel to the target, and the distance between them is the focal length of the metalens. The magnification imaging system is composed of metalens, a tube lens (Edmund, MT-40), and a CCD camera (Allied Visio, G032). The magnification imaging system is installed on a linear translation stage.